# RadioAstron: An Earth-Space Radio Interferometer with a 350,000 km Baseline

N. S. Kardashev, Astro Space Center of the P. N. Lebedev Physical Institute, Moscow, Russia; E-mail: nkardash@asc.rssi.ru

Y. Y. Kovalev, Astro Space Center of the P. N. Lebedev Physical Institute, Moscow, Russia; E-mail: yyk@asc.rssi.ru

K. I. Kellermann, National Radio Astronomy Observatory, 520 Edgemont Rd., Charlottesville, VA, 22901, USA; E-mail: kkellerm@nrao.edu

Abstract

RadioAstron is a Russian space based radio telescope with a ten meter dish in a highly elliptical orbit with an eight to nine day period. RadioAstron works together with Earth based radio telescopes to give interferometer baselines extending up to 350,000 km, more than an order of magnitude improvement over what is possible from earth based very long baseline interferometry. Operating in four frequency bands, 1.3, 6, 18, and 92 cm, the corresponding resolutions are 7, 35, 100, and 500 microarcseconds respectively in the four wavelength bands.

## 1. Resolution of Radio Telescopes

The first observations of cosmic radio emission in the 1930s by Karl Jansky [1,2,3] using a Bruce Array only a few wavelengths across at 15m wavelength had an angular resolution of a few tens of degrees. Later in the decade Grote Reber [4] used a 32 foot parabolic dish at 2 m wavelength, to improve the resolution to about 4 degrees. Since, the resolution of any instrument is limited by diffraction to an angle θ of the order of the wavelength of observation divide by the dimensions of the instrument, and because radio wavelengths are longer than light waves by a factor $\sim 10^5$, for many years it was thought that the resolution of radio telescopes would always be limited compared with that of optical telescopes.

However, for several reasons, it turns out that the reverse is true. First, due to their longer wavelength, it is much easier to build large diffraction limited telescopes at radio than at optical wavelength since the required mechanical tolerances are greatly relaxed at radio wavelengths.

Second, at optical wavelengths the path length fluctuations in the Earth's troposphere limit the resolution of optical observations by "seeing" to angles of the order of an arcsecond. Although, recent developments in adaptive optics using nearby stars or laser signals as a reference at infrared wavelength, and optical telescopes operating from space, such as the Hubble Space Telescope, are able to obtain resolutions somewhat better than 0.1 arcsec, at radio wavelengths, troposphere path length fluctuations are only comparable with the wavelength, and can easily be calibrated from observations of reference sources. For these reasons, the history of radio astronomy has been one of ever improving the angular resolution through a series of innovative technical developments.

The largest fully steerable radio telescopes are those in Effelesberg, Germany and in Green Bank, West Virginia in the USA, each with an effective diameter of 100 meters. Variations in mechanical tolerances due to wind, temperature variations across the structure and the effects of gravity as the structure is moved limit operation to wavelengths as short as about 1 cm to give an angular resolution of about 30 arcseconds.

## 2. Radio Interferometry

Unlike light waves, radio signals from one part of the telescope can be amplified, split, and compared coherently with signals from other parts of the instrument. Starting in the 1940s, radio astronomers began to use widely spaced interferometers of modest size parabolic antenna elements to give resolutions determined by the interferometer spacing and not by the dimensions of the individual antennas. Since each interferometer pair measures one Fourier component of the brightness distribution of the radio source, observations with multiple element arrays are used to reconstruct the two dimensional radio source structures [5]. In the typical radio interferometer or array a common local oscillator signal is sent to each of the distant elements where it is used to convert the received RF signal to a lower intermediate frequency which is then sent to a common point where it is correlated with the signals from the other antennas. One of the most powerful radio telescopes of this type is the recently upgraded Karl G. Jansky Very Large Array (VLA) located in central New Mexico. With twenty seven antenna elements, each of 25-m diameter, the resolution of the VLA at 1 cm wavelength in its largest configuration is about 0.1 arcseconds, comparable to that of the Hubble Space Telescope. See Figure 1.

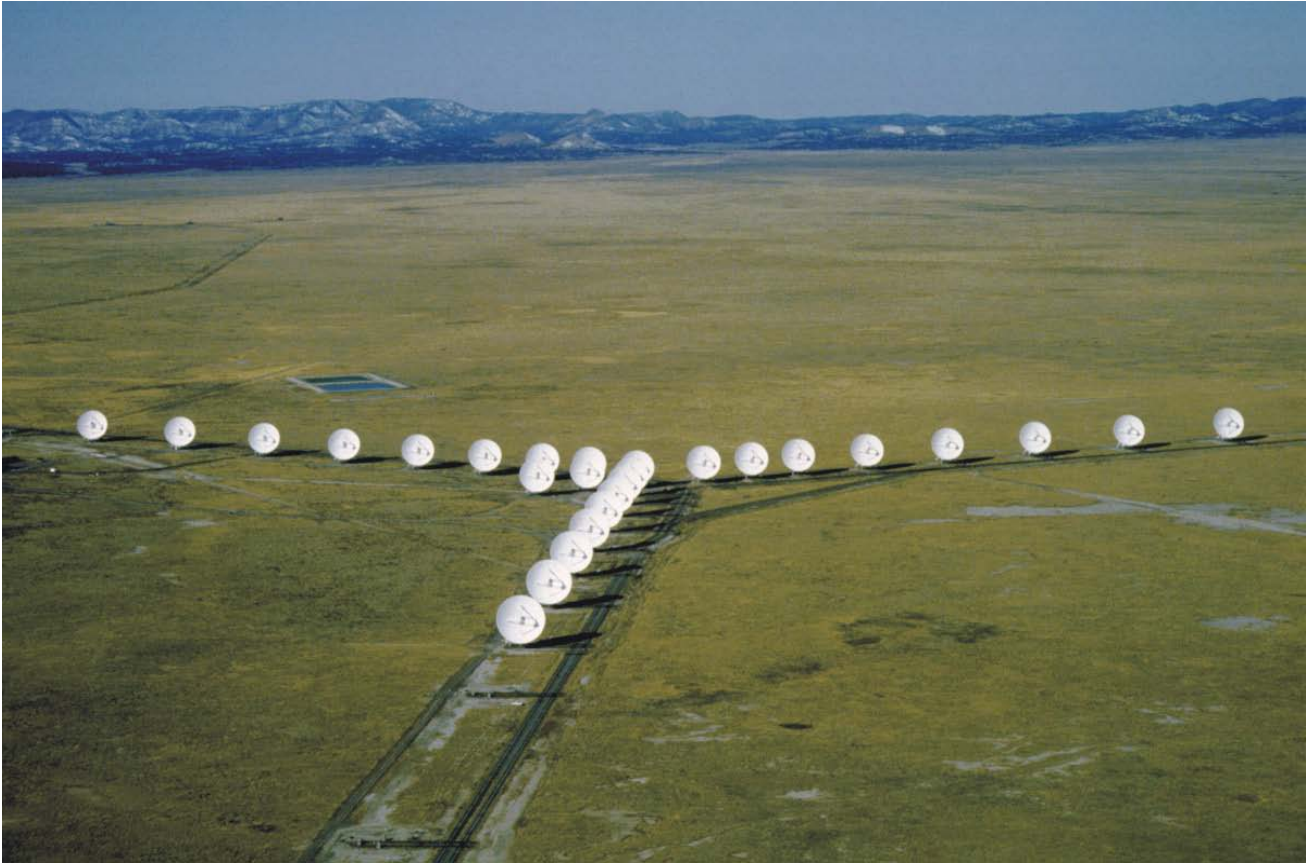

Figure 1. The Karl G. Jansky Very Large Array (VLA) located in central New Mexico includes twenty seven 25 meter antennas operating from a few hundred MHz to 50 GHz. Each antenna moves on railroad tracts along three arms to cover configurations ranging from 1 to 36 km. The angular resolution at the shortest wavelength and in the largest configuration is 0.04 arcseconds, about that of the Hubble Space Telescope.

Although in principle, the dimensions of radio interferometers can be extended without limit, practical considerations of routing the local oscillator transmission and broadband IF transmission lines have limited conventional interferometers and arrays to dimensions of the order of a few tens of kilometers. It is also possible to use radio links to distribute the IF and local oscillator signals within a multi-element array, but the bandwidths are limited by available spectrum allocations. More recently, the global fiber optic network has become an effective means to join antennas spaced hundreds and even thousands of miles apart. However, unless subsidized, the cost required to transmit IF bandwidths of hundreds of megahertz large data rates are prohibitively expensive.

## 3. Independent Oscillator Remote Recording Interferometry

To avoid these complexities, radio astronomers routinely implement very long baseline interferometers (VLBI) using independent local oscillators stabilized by hydrogen maser or other atomic frequency standards [6]. Until a few years ago, the data was recorded on magnetic tape at data rates up to 256 Mbps. However, the often unreliable and logistically difficult tapes have now been replaced with conventional computer disc drives. Compatible independent oscillator disc recording VLBI systems are now in routine use in the U.S., Europe, Asia, and Australia. Separate networks of antennas in the U.S., Europe, Russia, and East Asia routinely operate with nominal recording rates up to 512-1024 Mbps, and prototype systems are being tested at 2 and 4 Gbps record rates.

## 4. Space VLBI: *TDRSS* and *HALCA*

Interferometer baselines for these earth based antennas are of course limited to some fraction of an Earth diameter. Higher angular resolution can only be obtained by placing one end of the radio interferometer in space. In 1986 a team of scientists from the U.S., Japan, and Australia used an antenna on board the NASA Tracking and Data Relay Satellite at 2.3 and 15 GHz together with ground based antennas of the NASA Deep Space Tracking Network to demonstrate the feasibility of radio interferometry using an orbiting space craft with projected interferometer baselines up to 2 earth diameters [7]. In 1997, Japanese radio astronomers placed an 8-m diameter antenna aboard the HALCA (Highly Advanced Laboratory for Communications and Astronomy) spacecraft in low earth orbit. Unfortunately, the 22 GHz (1.3 cm) radiometer failed on launch so that observations with the remaining 5 GHz (6cm) and 1.6 GHz (18 cm) systems had a resolution only comparable to that achieved with conventional ground based systems operating at shorter wavelengths [8].

## 5. RadioAstron: Development and Specifications

On July 18, 1981 Roald Sagdeev then Director of the Soviet Cosmic Research Institut in Moscow, authorized the development of the *Spectrum-R* space interferometer mission, also known as *RadioAstron* [9]. RadioAstron was to be one of three missions; the others being Spectrum-X-Gamma and Spectrum-UV to operate at X/$\gamma$-rays and ultra-violet wavelengths respectively. However, the challenging technical goals of the three missions, combined with the political and economic difficulties following the fall of Soviet Union resulted in lengthy delays in the completion of RadioAstron which later was transferred to the Astro Space Center (ASC) of the P. N. Lebedev Physical Institute of the Russian Academy of Science. The Spectrum-X-Gamma and Spectrum -UV are currently under construction, the launch of the former is planned for 2014. The location of the major substructures is illustrated by the schematic diagram of the spacecraft shown in Figure 2. The assembled dish structure is shown in Figure 3 as seen by the visit of the RadioAstron International Steering Committee in 2008.

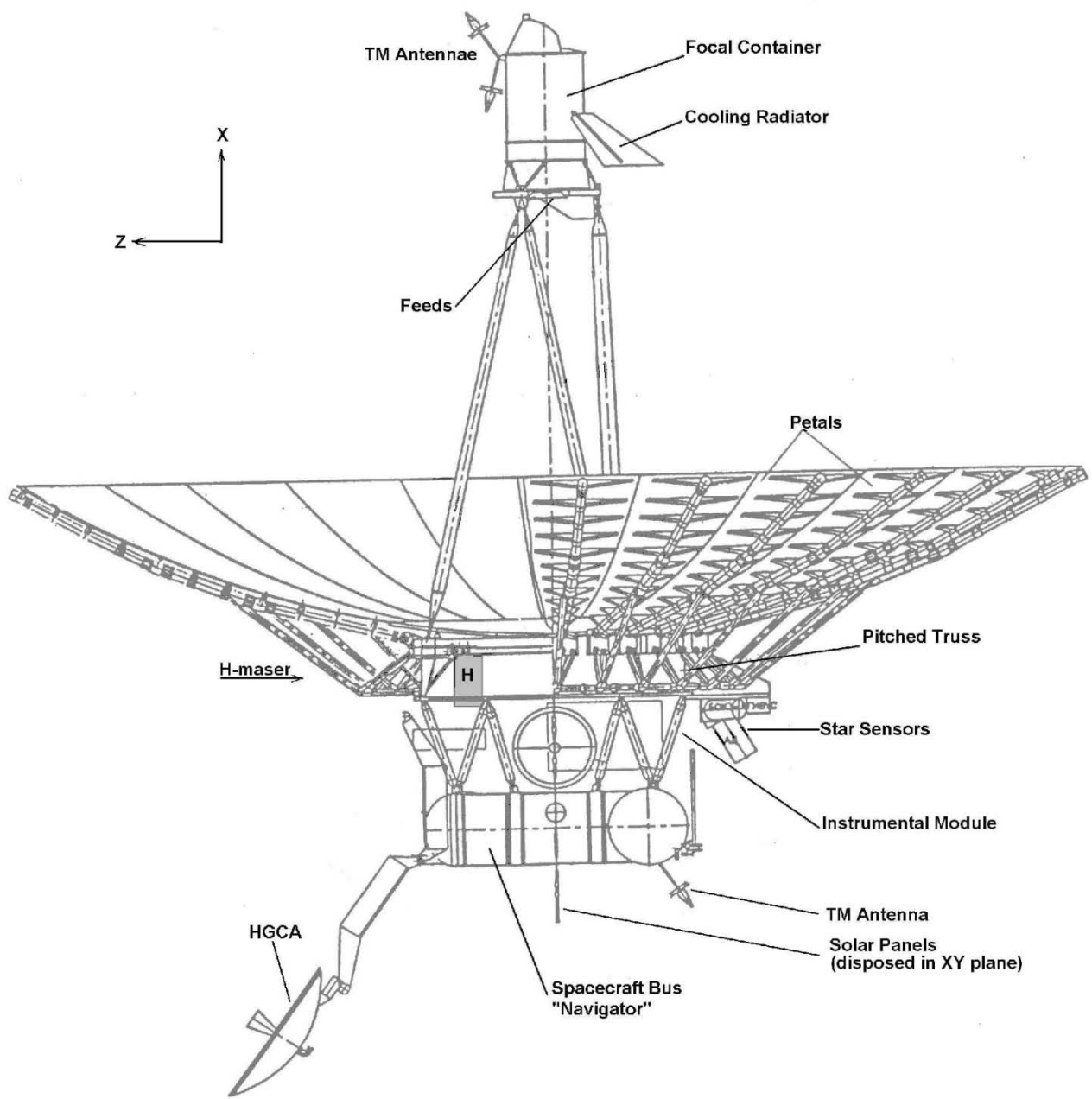

Figure 2. Schematic diagram of the RadioAstron Space Radio Observatory showing location of the Navigator service bus, the Hydrogen maser frequency standards, the High Gain Communications Antenna (HGCA) and other subsystems. The solar panels are oriented orthogonal to the diagram.

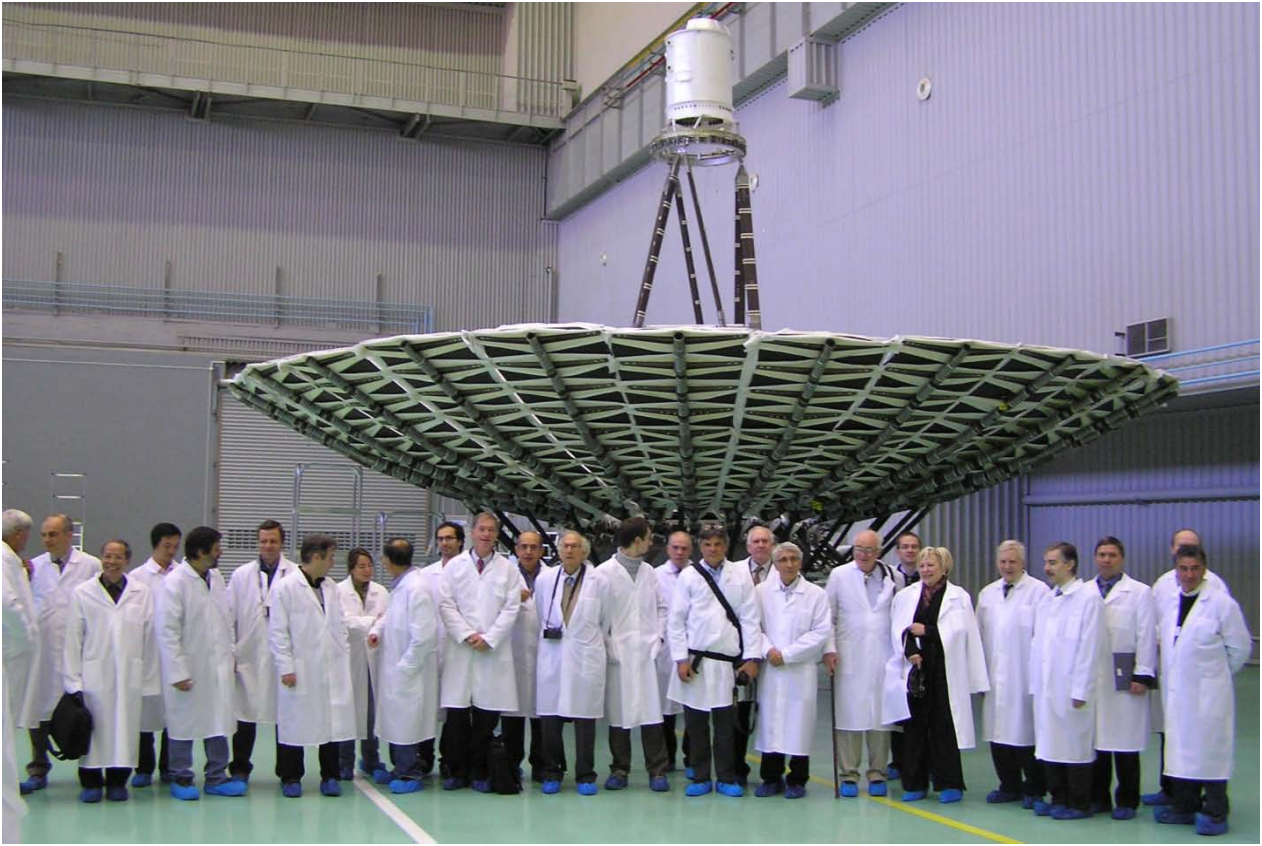

Figure 3. The 10 m space radio telescope as seen by the Radio Astron International Steering Committee during their visit to the Lavochkin Association construction facility in October 2008.

After 30 years of development of this ambitious Space VLBI project by scientists and engineers at the ASC, the 3660 kg RadioAstron spacecraft was finally launched from the Baikonour Cosmodrome on July 18, 2011, and a day later it was placed in a highly elliptical orbit extending out beyond 300,000 km with an eight to ten day period.

Table I. Main characteristics of the four RadioAstron frequency bands

| Frequency (GHz) | Tsys (K) | θ (μarcsec) | σ (mJy) |
|---|---|---|---|
| 0.33 | 200 | 530 | 16 |
| 1.66 | 45 | 100 | 2 |
| 4.6 | 130 | 35 | 4 |
| 22 | 77 | 7 | 9 |

The 10-m space RadioAstron radio telescope was fabricated from 27 carbon fiber panels and has a F/D ratio of 0.43. RadioAstron is equipped with feeds and receivers for four frequency bands as shown in Table I which gives in column 1 the frequency band; in column 2 the typical system temperature, Tsys, in column 3 the maximum resolution, and the rms noise in a typical coherence time of 5 minutes for 32 MHz bandwidth (16 MHz for 92 cm) when joined with the NRAO 100-m radio telescope in Green Bank, West Virginia, USA. The front side of the unfurled dish structure is seen in Figure 4 while Figure 5 shows the dish with the 27 panels folded and ready to be place in the rocket. Figure 6 shows the folded antenna installed in the Zenit rocket. Figure 7 conveys an artist's impression of the RadioAstron spacecraft in orbit with the antenna and solar panels unfurled.

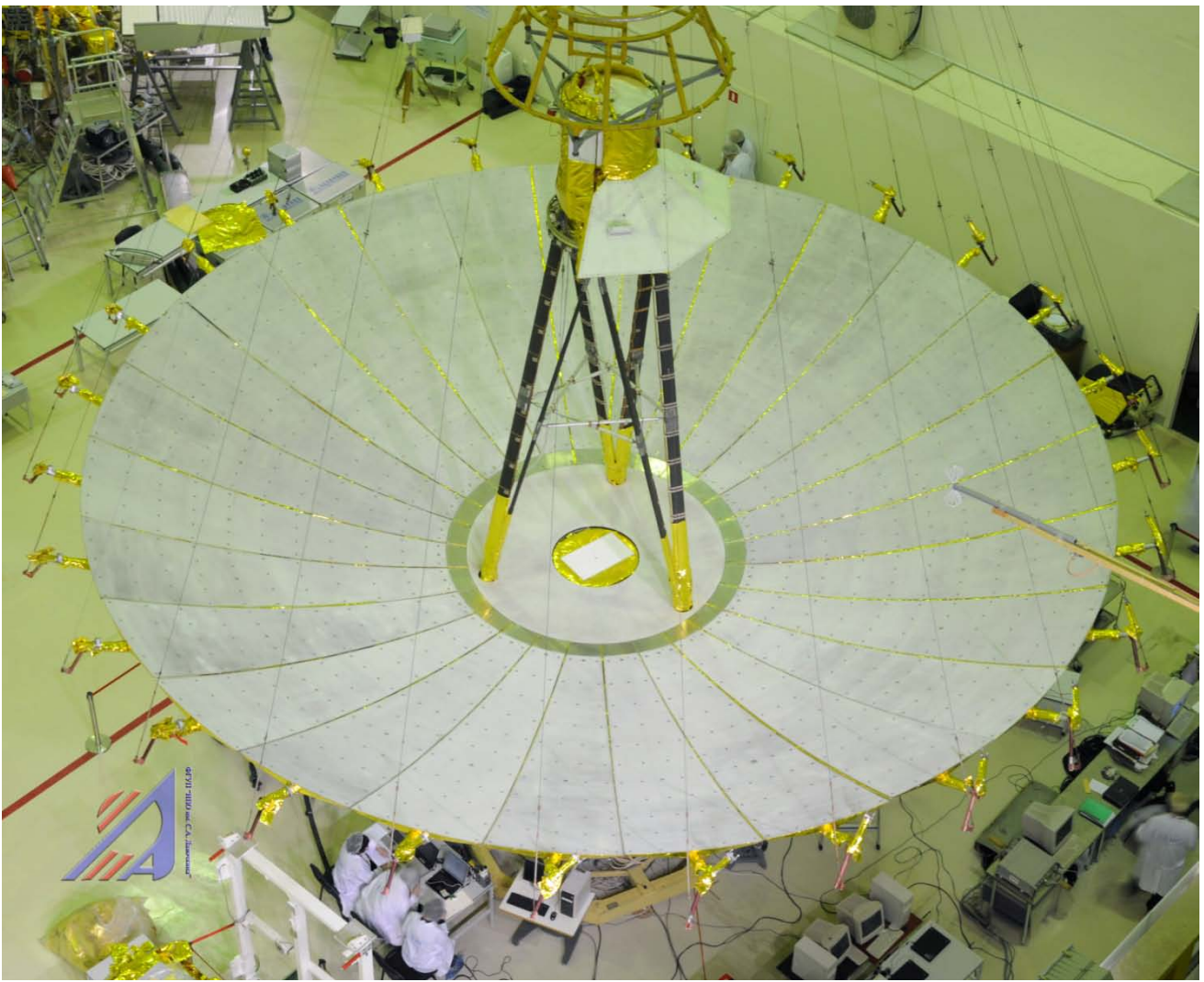

Figure 4. The Space Radio Telescope with the petals unfurled in the Lavochkin laboratory.

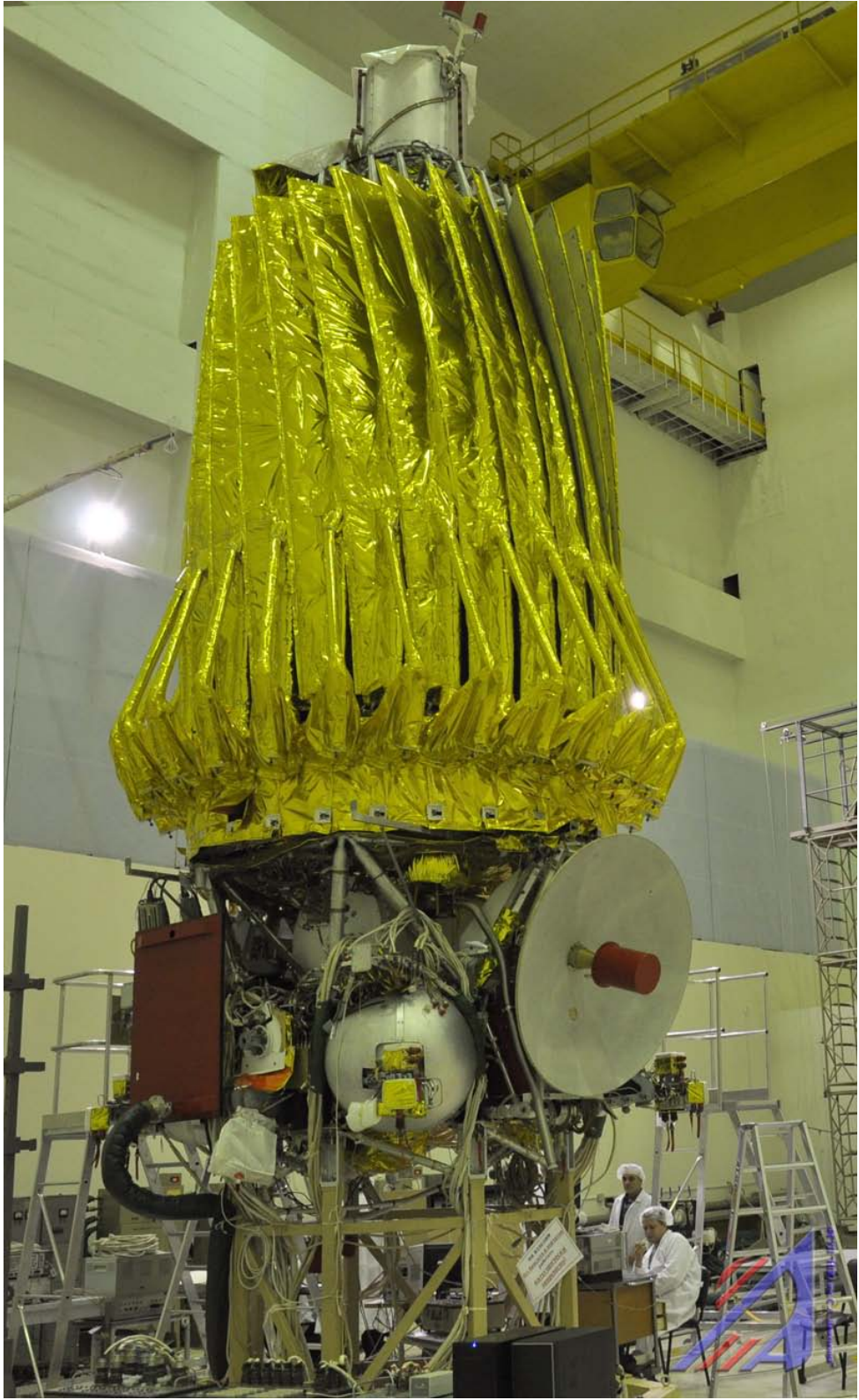

Figure 5. The 10 m RadioAstron dish shown with the 27 petals folded and ready for launch in the Zenit rocket.

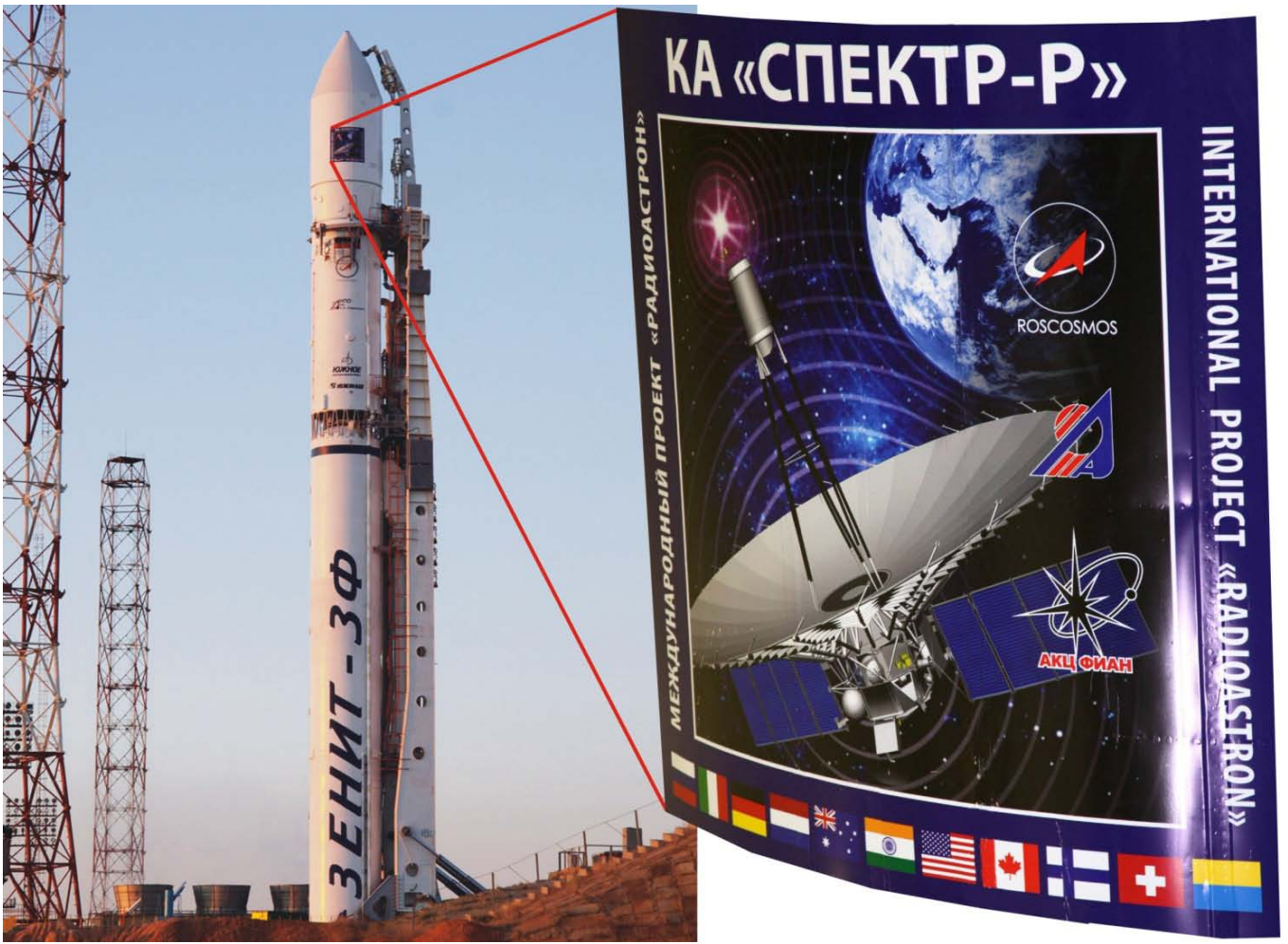

Figure 6. RadioAstron sitting on the Zenit rocket at the Baikonur launch site ready for launch. Insert shows the imbedded plaque displaying the international nature of RadioAstron.

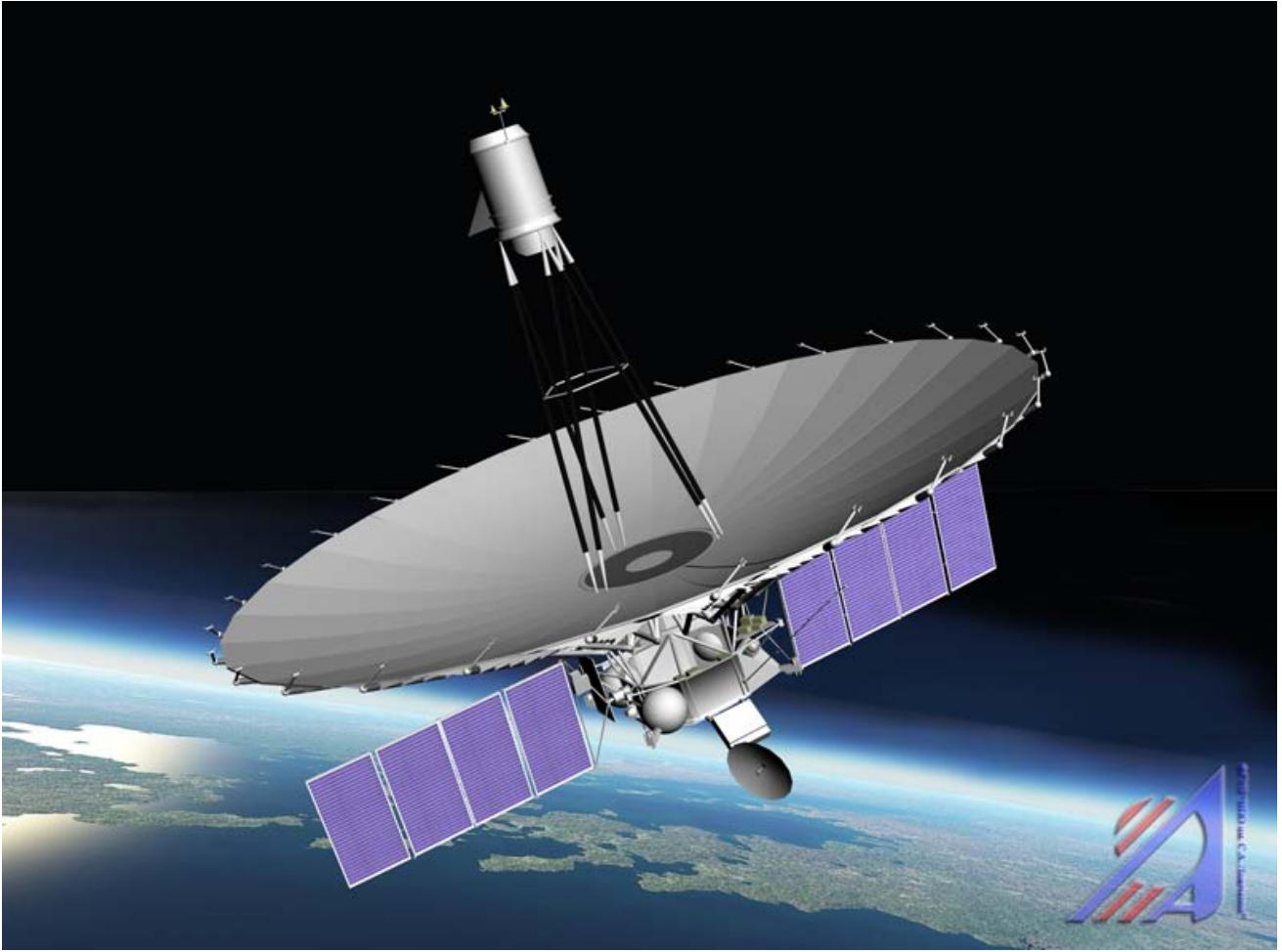

Figure 7. Artist's conception of the RadioAstron spacecraft.

The RadioAstron mission has enjoyed widespread international participation. The space radio telescope, the spacecraft bus, and instrumentation were designed and developed by the Astro Space Center, the Moscow S. A. Lavoshkin Federal Research and Production Association Industries, and the Russian Space Agency, Roscosmos, while the specialized radio interferometry instrumentation was developed at the Astro Space Center. The low noise amplifier for the 330 MHz (92 cm) radiometer was built by the NCRA in India and the 1.6 GHz (18 cm) receiver was manufactured by CSIRO in Australia. A 5 GHz (6 cm) receiver was constructed in the Netherlands on behalf of a consortium of European radio observatories. The 6 cm LNA was provided by the MPIfR in Germany, and the European Space Agency conducted thermal tests of the antenna panels in their vacuum chamber located in the Netherlands. The 22 GHz (1.3 cm) receiver was initially designed and built by the Helsinki Technical University. Due to the delays in the RadioAstron launch, both the original 5 and 22 GHz receivers were considered to have exceeded their shelf life and were replaced by newer units built by Russian industry. To obtain enhanced sensitivity and cover a wide frequency range from 18 to 25 GHz, the 22 GHz system uses low noise HEMT amplifiers constructed by the U.S. National Radio Astronomy Observatory. These are the same amplifier types as used for IF amplifiers in the WMAP spacecraft used to map the cosmic ray anisotropies. Each RadioAstron receiver operates in two (USB and LSB) 16 MHz wide channels (4 MHz wide at 330 MHz) in each of two orthogonal circular polarizations.

After conversion to baseband each 16 MHz intermediate frequency signal is digitized using one bit Nyquest sampling, and the 128 Mbps digital data stream is sent to the ground over a 15 GHz link. At the ground, the 128 Mbps data is recorded on RadioAstron Data Recorder discs [10]. The data is then played back and sent over fiber to Moscow or to the Max-Planck-Institute in Bonn, Germany, where they are correlated with data recorded at various ground radio telescopes throughout Russia, Ukraine, Europe, China, the United States, Japan, Australia, South Africa, and India. Correlation of the incoming data streams is performed in Moscow using a high performance computer cluster and specialized RadioAstron software correlator developed by the ASC team [10] and in Germany using a modified version of the standard DiFX software correlator [11].

The RadioAstron spacecraft is equipped with two hydrogen masers, manufactured by the Russian company Vremya-Ch. The masers are used to stabilize the on board local oscillator system by generating a 5 MHz reference signal used to control a frequency synthesizer which provides the independent stable local oscillator signal. As a back-up, in case of failure of the masers, a closed loop system operates at 7.2/8.4 GHz which can synchronize the RadioAstron local oscillator with one at the ground tracking station.

Ground tracking support is currently provided using the 22-m antenna located at the radio observatory near Puschino, outside of Moscow. Since the spacecraft is not always visible from Puschino, an additional tracking station is under development at the U.S. National Radio Astronomy Observatory in Green Bank, as well as one in South Africa. The South African tracking station will provide critical tracking support when the spacecraft is near perigee in the Southern Hemisphere. These external tracking stations will use the same instrumentation as at Puschino, thus insuring the uniformity of the data to facilitate the correlation. Precise orbit determination needed for radio interferometry is made utilizing five different methods – conventional radio delay and Doppler measurements, laser ranging, optical observations of the space craft sky position, VLBI tracking.

Data recorded simultaneously at the various tracking stations on the Earth and from RadioAstron are being used to reconstruct crude but extremely high resolution images of celestial radio sources. Because of the very elliptical orbit of RadioAstron, the resolution of the earth-space interferometer is essentially one dimensional. Precession of the orbit resulting from gravitational perturbations by the Moon, will give some degree of two-dimensional coverage, although changes in the radio source structure during this period will make the detailed interpretation of the interferometer data more difficult.

## 6. RadioAstron Scientific Goals

Probable targets of Radioastron studies include pulsars, blazars, and cosmic masers. Previous ground based VLBI observations of blazars suggest the presence of structure on angular scales as small as 50 microarcseconds [12]. These expected small scale structures are supported by observations of the radio spectra and rapid time variability. Of particular interest are the so called blazars, which are quasars whose relativistic outflow is directed toward the earth. Because of the enhancement of the radio synchrotron emission which is beamed along the direction of motion due to relativistic boosting, the apparent brightness of blazars can appear boosted by factors of thousands. Also, since the radio emitting plasma is thought to be moving at nearly the speed of light along the line of sight, the radiating source is nearly keeping up with its own radiation, giving the appearance of faster than light motion. The high resolution observations made possible with RadioAstron, will allow unprecedentedly detailed observations of blazars, reaching an order of magnitude closer to the super massive black holes thought to lie at the base of the relativistic jets.

Also, clouds of hydroxyl (OH) ions and $H_2O$ molecular gas are found in the regions surrounding highly evolved stars as well as regions where new stars are being formed. Excited by uv radiation from the associated star, these clouds can act as cosmic masers giving intense rapidly variable radio emission from very small regions.

Pulsar radio emission is formed within the .highly organized strong magnetic fields surrounding rapidly rotating neutron starts. The pulsar radiation comes from such small regions that will remain unresolved, even by RadioAstron. However, the RadioAstron observations, especially at the longer wavelengths, will study the very small scale structures in the intervening intergalactic medium which scatters the radiation resulting in apparent angular dimensions greater than the intrinsic values.

Interferometer baselines between RadioAstron and ground based radio telescopes have an angular resolution more than an order of magnitude better than used in any previous astronomical observation. Specifically, at 22 GHz, the longest baselines are more than $2 \times 10^9$ wavelengths giving an angular resolution of only $7 \times 10^{-6}$ arcseconds to study the radio emission from quasars and cosmic masers. Early science observations have been in progress since February, 2012, and starting in mid 2013, access to RadioAstron and the supporting ground facilities will be open to peer review proposals from any scientist, independent of their institutional or national affiliation.

## 7. Early Results

Following the launch of the spacecraft in July, 2011, the first four months in orbit were spent in checking and calibration of the various mechanical and electronic components and in calibrating the pointing of the antenna using the Moon and other strong cosmic radio sources. These observations confirmed the performance of the four radiometers each of which had a measured system temperature close to the design value.

Small uncertainties in the position and motion of the spacecraft at any time result in corresponding uncertainties in the interferometer fringe rate and in the path length delay between the RadioAstron and the ground antennas. A series of fringe-finding observations at each of the four observing frequencies was begun in November, 2011, to verify the performance in each of the four frequency bands and to determine the corresponding residual fringe rate and delay for a variety of cosmic sources. Since RadioAstron will observe primarily in a previously unexplored range of angular resolution, the early fringe finding observations were mostly made using strong sources with projected interferometer spacing that are comparable with previous earth based observations having sufficient fringe amplitude for detection with RadioAstron. These observations resulted in the successful detection of fringes in all four RadioAstron frequency bands. In three of them, 18, 6, and 1.3 cm first fringes were found on a quasar while the 0.3 GHz band was successfully tested using a bright pulsar on a very long projected baseline. The detection of interference fringes is illustrated in Figure 8 which shows the observed fringe amplitude as a function of residual fringe rate and delay compared with the values corresponding to the space to ground interferometer baseline calculated from the spacecraft orbit parameters.

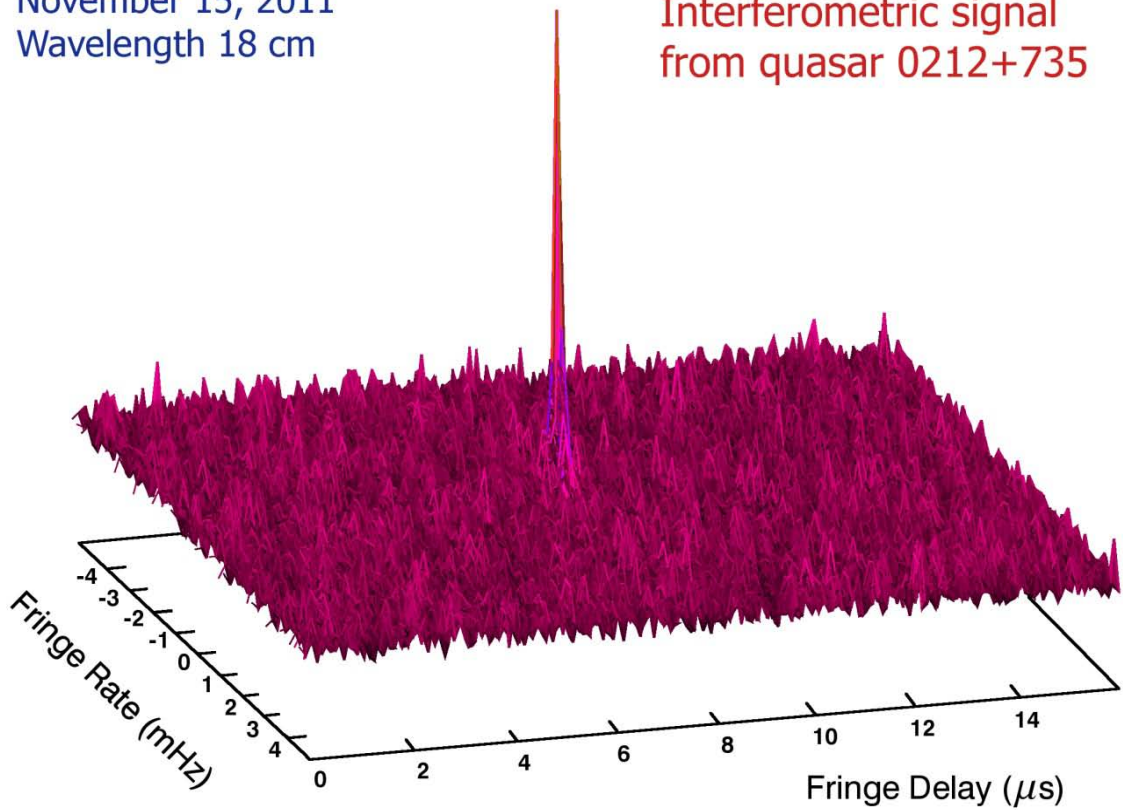

Figure 8. The first fringes found by the space-VLB interferometer RadioAstron. Interference signal from the quasar 0212+735 on a 8,100 km projected baseline between RadioAstron and the 100-m MPIfR radio telescope near Effelsberg, Germany, observed on November 15, 2012. The wavelength of the observations was 18 cm. The plot shows the fringe amplitude versus residual delay and residual fringe rate in a single 16 MHz wide channel.

Regular scientific observations with RadioAstron are organized by three international RadioAstron early science program working groups being coordinated by Astro Space Center. They have been underway since February 2012 with 0.3 and 1.6 GHz observations of pulsars; 1.6 GHz observations of OH masers; 22 GHz observations $H_2O$ masers; as well as 1.6, 5, and 22 GHz observations of quasars. By the mid 2012 important results were achieved by all the groups with quasar detection up to projected interferometer baseline of 92,000 km (7.2 Earth diameters), pulsar detection at up to 220,000 km (about 20 Earth diameters) and water maser detection at 1.3 cm just over 1 Earth diameter.

**Acknowledgments.** The ambitious RadioAstron program owes its success to the dedicated scientists and engineers at the Astro Space Center and Lavochkin Association, many of whom have worked for decades to bring the mission to fruition, as well as to the international observatories (Kvazar network, Russia; Evpatoria, Ukraine; Effelsberg, MPIfR, Germany; Medicina and Noto, Italy; Yebes, Spain; Westerbork, the Netherlands; other EVN telescopes; Arecibo and GBT, USA; Usuda, Japan; LBA telescopes and Tidbinbilla, Australia; etc.) and colleagues who have supported the development and early operations. This paper has used material from the RadioAstron user hand book (http://www.asc.rssi.ru/radioastron/documents/rauh/en/rauh.pdf) and RadioAstron Newsletter (http://www.asc.rssi.ru/radioastron/news/news.html). More detailed information on RadioAstron can be found on the RadioAstron web site, http://www.radioastron.ru. The National Radio Astronomy Observatory is a facility of the National Science Foundation operated under cooperative agreement by Associated Universities, Inc. Y.Y.K. was supported in part by the Dynasty Foundation.